\documentclass[5p,sort&compress]{elsarticle}

\usepackage{comment}
\usepackage[squaren]{SIunits}
\usepackage{bm,amsmath,amsfonts,braket,todonotes,subfig,multirow,dcolumn}

\renewcommand{\vec}[1]{\ensuremath{\bm{#1}}}
\newcommand{\op}[1]{\ensuremath{\hat{#1}}}
\newcommand{\vecop}[1]{\ensuremath{\op{\vec{#1}}}}
\newcommand{\mechanical}[1]{\ensuremath{\mathfrak{#1}}}

\newcommand{\expectation}[1]{\left\langle #1 \right\rangle}

\newcommand{\ii}{\ensuremath{\mathrm{i}}}
\newcommand{\sgn}{\operatorname{sgn}}

\begin{document}
\title{Peculiar Rotation of Electron Vortex Beams}

\author[ifp,ustem]{T. Schachinger\corref{cor1}}
\ead{thomas.schachinger@tuwien.ac.at}
\author[ifp,ustem]{S. L\"offler}
\author[ustem]{M. St\"oger-Pollach}
\author[ifp,ecp]{P. Schattschneider}
\cortext[cor1]{Corresponding author}

\address[ifp]{Institute of Solid State Physics, Vienna University of Technology, Wiedner~Hauptstra{\ss}e~8-10, 1040 Vienna, Austria}
\address[ustem]{University Service Centre for Transmission Electron Microscopy,
Vienna University of Technology, Wiedner Hauptstra{\ss}e 8-10, 1040 Wien, Austria}
\address[ecp]{LMSSMat (CNRS UMR 8579)m Ecole Centrale Paris, F-92295 Ch\^atenay-Malabry, France}


\begin{abstract}
Standard electron optics predicts Larmor image rotation in the magnetic lens field of a TEM. Introducing the possibility to produce electron vortex beams with quantized orbital angular momentum brought up the question of their rotational dynamics in the presence of a magnetic field. Recently, it has been shown that electron vortex beams can be prepared as free electron Landau states showing peculiar rotational dynamics, including no and cyclotron (double-Larmor) rotation. Additionally very fast Gouy rotation of electron vortex beams has been observed.
In this work a model is developed which reveals that the rotational dynamics of electron vortices are a combination of slow Larmor and fast Gouy rotations and that the Landau states naturally occur in the transition region in between the two regimes.
This more general picture is confirmed by experimental data showing an extended set of peculiar rotations, including no, cyclotron, Larmor and rapid Gouy rotations all present in one single convergent electron vortex beam.

\end{abstract}

\maketitle
\section{Introduction}
Vortex beams are  characterized by a spiraling wavefront with a phase singularity at the center. First signs of this phenomenon have already been observed in the 1950's~\cite{Findlay1951}, but possibly due to the lack of an accompanying theory, which was delivered nearly a quarter of a century later by Nye and Berry~\cite{NyeBerry1974}, it took until the 1990's for the first intentional experimental realization with light waves~\cite{Bazhenov1990,Allen1992}. 
Today, there are many applications of optical vortices ranging from tweezers exerting a torque~\cite{HePRL1995}, over optical micromotors~\cite{LuoAPL2000}, cooling mechanisms~\cite{KuppensPRA1998}, toroidal Bose-Einstein condensates~\cite{TsurumiJPSJap2001},  communication through turbulent air~\cite{Krenn2014} to exoplanet detection~\cite{SerabynNature2010}.
A similar story holds true for vortex \emph{matter waves}. In the early 1970's Beck, Mills and Munro produced helical spiraling electron beams~\cite{Beck1971, Mills1971, Munro1971, Beck1973} using a special magnetic field configuration. However, the significance of that discovery was most likely underestimated until recently, when Bliokh {\it et al.} described phase vortices in electron wave packets~\cite{Bliokh2007}. This paved the way for the first experimental realization of electron vortex beams (EVBs) in a transmission electron microscope (TEM)~\cite{UchidaNature2010}. Shortly after, the holographic mask technique was introduced for routinely producing electrons with quantized orbital angular momentum (OAM) in the TEM~\cite{VerbeeckNature2010}. Up to now, several methods for producing vortex beams with higher orders or higher brilliance have been published~\cite{McMorranScience2011, Schattschneider2012, Beche2014, Grillo2014}, reflecting the vital interest in shaping the electron wavefront.
 
Owing to their short wavelength, fast EVBs ($\sim$~100 to 300~keV) can  be focused to atomic size~\cite{Schattschneider201221}. Another interesting aspect for future applications is their quantized orbital magnetic moment $\mu_B m$, which is -- at least for fast electrons, where spin-orbit coupling can be neglected -- independent of spin polarization. This allows the creation of electron beams carrying magnetic moment even without spin-polarization. Both features make them attractive as a novel probe in solid state physics for mapping, e.g. magnetic properties~\cite{Rusz2013, Schattschneider2014a} on the atomic level. In addition, EVBs could be used to probe Landau states (LS)~\cite{Bliokh2012,Schattschneider2014} and have already been shown to be a promising candidate for manipulating nanoparticles~\cite{Verbeeck2013}.

The theory of propagating EVBs has been developed in a series of publications~\cite{Bliokh2007, Schattschneider2011,BliokhPRL2011, Bliokh2012,LoefflerACA2012,LubkPRA2013}. The most intriguing prediction is the peculiar rotation mechanism of  vortex beams in a magnetic field. It was shown that exact solutions of the paraxial Schr\"odinger equation in a homogeneous magnetic field --- non-diffracting Laguerre-Gaussian modes also known as LS --- acquire a phase upon propagation along the $z$ axis~\cite{Bliokh2012}. This phase shift causes a quantized rotation of coherent superpositions of LG modes, falling in one of three possible groups showing either cyclotron, Larmor, or zero frequency, depending on the topological charges involved.

This surprising prediction was confirmed experimentally with beams closely resembling non-diffracting solutions of the Schr\"odinger equation in a homogeneous magnetic field~\cite{Schattschneider2014}. Contrary to this quantized rotation, rapid Gouy rotation of EVBs close to the focus has been observed~\cite{GuzzinatiPRL}. Both experiments seem to contradict each other, as well as the standard theory of electron movement in a TEM that predicts Larmor rotation of paraxial ray pencils between object and image in a round magnetic lens~\cite{Glaser1952}.  These facts call for a re-evaluation of the rotation dynamics of convergent and divergent electron beams in the TEM, including beams with non-vanishing topological charge.

Here, we give a quantum description of  {\it diffracting} electron vortices based upon radius $r$, angular momentum $m\hbar$, and their time evolution. It is found that the rotation dynamics are a function of the OAM and the expectation value $\expectation{\hat r^{-2}}$ (the second moment of the inverse beam radius). This finding is supported by experiments tracking the rotation of convergent electron waves with topological charge in the lens field. 
It is shown that the whole range of rotation dynamics, namely 'classical' Larmor rotation (LR), rapid Gouy rotation, cyclotron (double-Larmor) and zero rotation occurs in one beam, and can be described within a uniform picture.

\section{Theory}
We discuss the  dynamics of electron vortex beams (and their superpositions) in a constant homogeneous magnetic field pointing in $z$-direction, $\vec{B}$. Such a field gives rise to the vector potential
\begin{equation}
	\vecop{A} = -\frac{\vecop{r} \times \vec{B}}{2}
	\label{eq:vector-potential}
\end{equation}
in the Coulomb gauge. The Hamiltonian of the system takes the form
\begin{equation}
	\op{H} = \frac{\vecop{\mechanical{p}}^2}{2m_e} = \frac{(\vecop{p} - e\vecop{A})^2}{2m_e},
	\label{eq:Hamiltonian}
\end{equation}
where  $\vecop{\mechanical{p}} = \vecop{p} - e\vecop{A}$ is the observable {\it kinetical} or {\it covariant} momentum operator, whereas the canonical momentum operator $\vecop{p}$ is gauge dependent and not an observable.
 
\subsection{Rotation dynamics}
\label{sec:RotDynamics}

The rotation dynamics of EVBs in a magnetic field is intimately connected to the concept of Bohmian trajectories. They can be interpreted as streamlines of the quantum mechanical particle current density \cite{Bliokh2012}
\begin{eqnarray}
\vec{j}(\vec{r})&=&\frac{1}{m_e} \Re[\psi^*(\vec{r}) \mechanical{\vec{p}}(\vec{r}) \, \psi(\vec{r})]= \nonumber \\
&=&-\frac{1}{m_e} \Re[\psi^*(\vec{r}) (i \hbar \nabla+e \vec{A}(\vec{r})) \, \psi(\vec{r})].
\label{eq:j}
\end{eqnarray}
 In the present context, we want to calculate the angular velocity of a quantum fluid following such streamlines.
 In cylindrical coordinates $\vec r=(r,\varphi,z)$ centered at the optical axis, the azimuthal velocity on a streamline is
\begin{equation}
v_\varphi(\vec{r})=j_\varphi(\vec{r})/|\psi(\vec{r})|^2.
\end{equation}
For a constant magnetic field in $z$-direction $B_z$ we have
$A_r(\vec{r})=A_z(\vec{r})=0$ and $A_\varphi(\vec{r})=r B_z/2.$
Since $ \nabla_\varphi =r^{-1} \partial_\varphi $, the angular velocity is
\begin{equation}
\omega(\vec{r})=\frac{v_\varphi(\vec{r})}{r}=\frac{\hbar}{m_e}\Im\left[\frac{\psi^*(\vec{r}) r^{-1} \partial_\varphi \, \psi(\vec{r})}{r \, \psi^*(\vec{r}) \psi(\vec{r})}\right]-\frac{e B_z}{2 m_e}.
\label{e6}
\end{equation}
The rotation dynamics  is experimentally accessible through the expectation value 
\begin{equation}
\expectation{\omega(z)}=\int{\psi^*(\vec{r})\, \omega(\vec{r}) \, \psi(\vec{r}) \,rdrd\varphi}.
\label{e7}
\end{equation}
Since in cylindrical coordinates the OAM operator is 
\begin{equation}
\op{L}_z = i \hbar \partial_\varphi
\label{e8}
\end{equation}
and $[\op{L}_z,\op{r}]=0$, we find from Eqs.~\ref{e6}--\ref{e8}
\begin{equation}
\expectation{\omega(z)}=\frac{1}{m_e}  \expectation{\op{r}^{-2}\op{L}_z}+ \sigma \Omega
\label{omega}
\end{equation}
where we have introduced the Larmor frequency $\Omega = |e B_z / 2 m_e|$. 
$\sigma=\sgn B_z=\pm1$ designates the direction of the axial magnetic field.
Eq.~\ref{omega} is one of our main results. It serves as basis for the following 
study of the rotation dynamics.

The expectation value of the angular velocity Eq.~\ref{omega} is conveniently obtained by decomposing the wave function $\psi(\vec{r})$ into normalized orthogonal eigenfunctions of $\op{L}_z$:
\begin{equation}
\psi(\vec{r})=\sum_m \psi_m(r,z)\,e^{\ii m \varphi}.
\end{equation}
Since $\op{L}_z \psi_m(r,z)\,e^{\ii m \varphi}=\hbar m \psi_m(r,z)\,e^{\ii m \varphi}$, it follows
\begin{equation}
\expectation{\op{r}^{-2}\op{L}_z}=\hbar \sum_m m \, \expectation { \op{r}^{-2}}_m 
\label{eq:ExpectValLzOverRsqrd}
\end{equation}
with
\begin{equation}
\expectation {\hat r^{-2}}_m =\int_0^\infty {\psi_m^\ast(\vec{r}) r^{-2} \psi_m(\vec{r}) \, r\, dr}.
\label{expectationRadius}
\end{equation}
Transforming to the dimensionless radial distance $\hat \xi= \hat r/w_B$, where
\begin{equation}
	w_B = \sqrt{\frac{2 \hbar}{|e B|}} = \sqrt{\frac{\hbar}{m_e \Omega}}
\end{equation}
is the magnetic beam waist, representing the radius that  encloses  one magnetic flux-quantum\footnote{The magnetic flux through a circle of radius $w_B$ is $w^2_B\, \pi \, B = h/e$.}, Eq.~\ref{omega} yields
\begin{eqnarray}
\expectation{ \omega(z)}&=&\frac{\hbar}{m_e w_B^2} \sum_m m \, \expectation { \hat \xi^{-2}}_m  +\sigma \Omega \nonumber \\
&=&\Omega\left(\sum_m m  \, \expectation { \hat \xi^{-2}}_m  +\sigma\right).
\label{eq:OmegaXi} 
\end{eqnarray}

Eq.~\ref{omega} and Eq.~\ref{eq:OmegaXi} show that vortices will rotate according to their radial extension. The wider the vortices are in a given observation plane, the smaller are their inverse square moments $\expectation{\hat \xi^{-2}}$. With that the first term in Eq.~\ref{eq:OmegaXi} gets negligible compared to the second, such that the rotation frequency asymptotically approaches the Larmor frequency. Depending on the magnetic field orientation $\sigma$, this rotation will be clockwise or anticlockwise, as long as the vortex radius is significantly larger than $w_B$. In this regime --- which will be referred to as LR-region throughout this manuscript --- the rotation is completely independent of the vortices' OAM.

When $\sum_m m  \, \expectation { \op {\xi}^{-2}}_m $ is close to $\pm1$, the EVBs approximate LS, see Sec.~\ref{section:DiffLGmodes}, showing no rotation for anti-parallel orientation of the OAM with respect to the magnetic field $B_z$ and cyclotron rotation (double-LR) for parallel orientation. This regime will be called LS-region. 

Obviously, for radial extensions smaller than $w_B$, which is of the order of  $w_B\sim\unit{25}\nano\meter$ for typical objective lens fields of the order of $B_z\sim\unit{2}\tesla$, the rotation frequency increases drastically. In this so-called Gouy-regime or Gouy-region, simulations show that vortices with a radial extension of $\sim\unit{1}\nano\meter$ and $m=1$ can rotate with $\sim 10^3 \, \Omega$, a rotation frequency that corresponds to the cyclotron frequency in fields of $\sim 1000$~T.

\subsection{Diffracting LG modes}
\label{section:DiffLGmodes}

To illustrate the properties of EVBs in the TEM, it is reasonable to consider a set of solutions of the paraxial Schr\"odinger equation, namely {\it diffracting LG (DLG) modes}~\cite{Schattschneider2014} given by:
\begin{eqnarray}
\psi_{m,n}(\vec{r}) &=& \sqrt{\frac{n!}{\pi (n+|m|)!}}\frac{1}{w(z)}\left(\frac{r}{w(z)}\right)^{|m|} \times \nonumber \\
 &&L_n^{|m|}\left(\frac{ r^2}{w(z)^2}\right) e^{-\frac{r^2}{2 w(z)^2}} e^{\ii\frac{k\,r^2}{2R(z)}}\times \nonumber \\
&& e^{-\ii(2n+|m|+1)\zeta(z)}  e^{\ii (m \varphi+k z)}.
\label{eq:DLG}
\end{eqnarray}
The parameter
\begin{equation}
w(z)=w_0\,\sqrt{1+\left(\frac{z}{z_R}\right)^2}
\label{eq:beamwaist}
\end{equation}
describes the transverse beam size evolution over $z$. $w_0$ represents the "beam waist radius" (of the $m=0$ beam)\footnote{$w(z)$ is not the maximum intensity radius of a vortex, which is given by $r_{max}(z,m)=w(z)\sqrt{|m|}$ for $n=0$.} at the focal plane $z=0$, $k$ stands for the forward wave vector and 
\begin{equation}
 z_R=\frac{\pi\,w_0^2}{\lambda}=k\,w^2_0,
\end{equation}
is the Rayleigh range, denoting the position where the illuminated area doubles and the acquired phase shift stemming from the Gouy phase given by $\zeta(z)=\arctan(z/z_R)$ reaches $\pi/4$. The curvature $R(z)$ of the LG mode is $R(z)=z\,(1+(z/z_R)^2)$. 
\begin{figure}[h]
\begin{flushleft}
\centering\includegraphics[width=90mm]{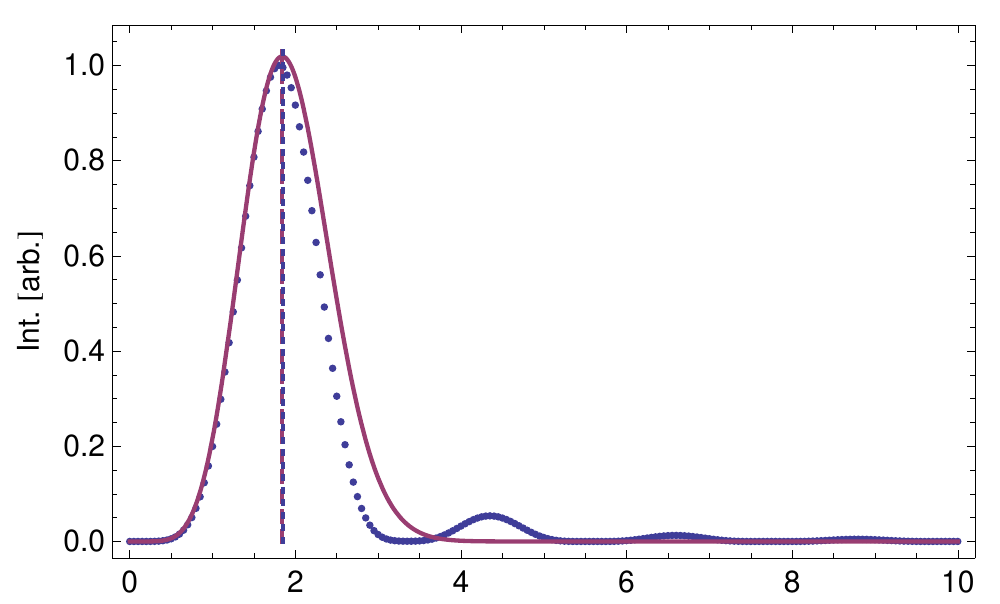}\\[-1pt]
\includegraphics[width=90mm]{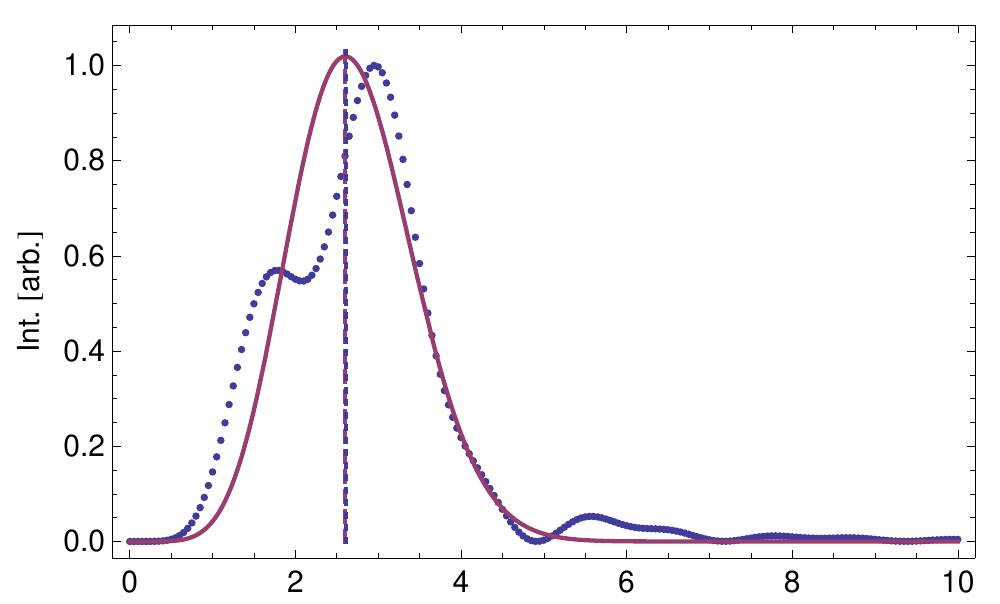}\\[-1pt]
\includegraphics[width=90mm]{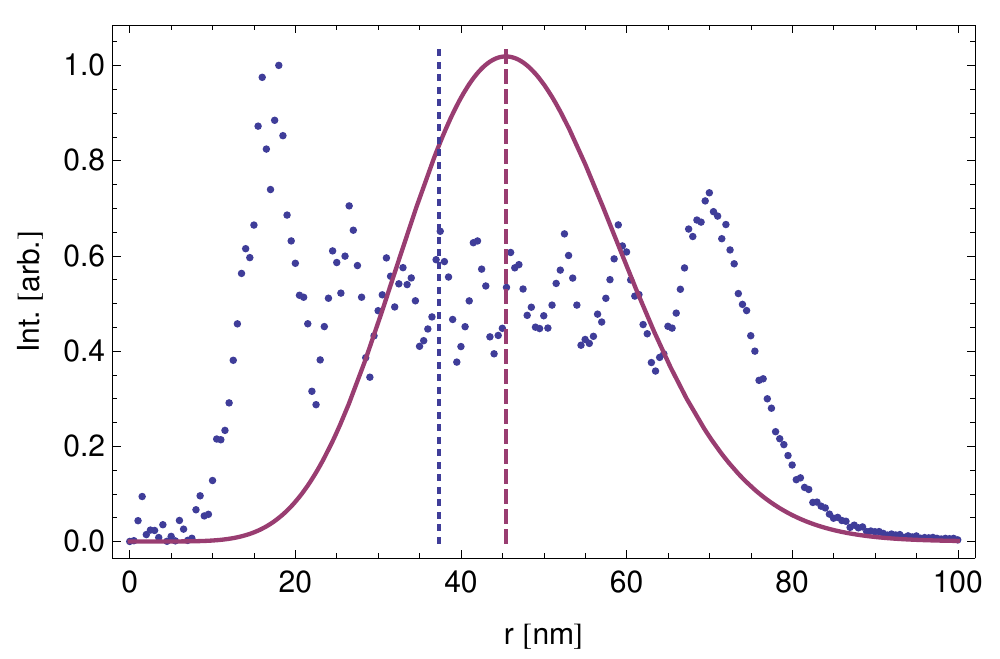}
\caption{Comparison between numerical simulated EVB (blue dots) and the analytical LG mode $\psi_{3,0}$ (purple solid line) in focus (uppermost), at $z_R$ (middle) and a defocus of $\unit{70}\micro\meter$ for an OAM $\hbar m=3\hbar$ EVB with $z_R=\unit{2.84}\micro\meter$. The blue dotted vertical lines represent the $\expectation{\op {\xi}^{-2}}^{-1/2}$ values of the numerical simulated EVB, whereas the purple dashed ones belong to the analytical LG modes.} 
\label{fig:LG_FFTcomparison}
\end{flushleft}
\end{figure}

To check the applicability of this approach for converging vortices in the TEM, Fig.~\ref{fig:LG_FFTcomparison} compares numerically simulated transverse beam profiles with the analytical approach Eq.~\ref{eq:DLG} of a $\psi_{3,0}$ mode for different $z$-values, see also Sec.~\ref{sec:Numerical simulation and error sources}. It illustrates that diffracting LG modes approximate the intensity distributions very well between the focal plane and the Rayleigh range $z_R$.
For higher $z$-values, the radial profiles deviate more and more from the LG modes, but as to the rotation dynamics, $\expectation{\op {\xi}^{-2}}_m=w_B^2/(|m|\,w(z)^2)$ is still close to that of the numerical simulated radial profiles, see Fig.~\ref{fig:LG_FFTcomparison}.

Now, the expectation value  $\expectation{\omega(z)}$ in Eq.~\ref{eq:OmegaXi} for a single mode results in 
\begin{equation}
\expectation {\omega(z)}=\Omega\left(\frac{m}{|m|}\left(\frac{w_B}{w(z)}\right)^2+\sigma\right),
\label{eq:analyticomega}
\end{equation} 
or with Eq.~\ref{eq:beamwaist} 
\begin{equation}
\expectation {\omega(z)}=\Omega\,\left(\frac{m\,w_B^2}{|m|w_0^2}\frac{1}{1+\left(z/z_R\right)^2}+\sigma\right).
\label{eq:analyticomega1}
\end{equation} 
Eq.~\ref{eq:analyticomega} and~\ref{eq:analyticomega1} show the salient features of convergent electron beams, which have already been described in Sec.~\ref{sec:RotDynamics}, including LR ($w(z) \gg w_B$), no rotation ($w(z) = w_B$, $\sigma m <0$), double-LR ($w(z) = w_B$, $\sigma m > 0$), and fast Gouy phase dynamics  ($w(z) \ll w_B$). They ultimately depend on the relative strength of the two terms in Eq.~\ref{eq:analyticomega} and~\ref{eq:analyticomega1} and thus the radial extension of the EVB. The intriguing thing is that these rich rotational dynamics are all contained in one single beam, due to its converging character, see Figs.~\ref{fig:LG_FFTcomparison} and ~\ref{fig:ExpSetup}.  

For a certain defocus value $z=z_B$ where $w(z_B)=w_B$
we find
\begin{equation}
\expectation { \omega(z_B)}=\Omega(\sgn(m)+\sigma).
\end{equation}
The angular frequency becomes quantized, with only three possible absolute values: 0, $\Omega$, or $2\Omega$ for $\sigma m<0$, $\sigma m=0$, $\sigma m >0$. 
This surprising result was derived in~\cite{Bliokh2012} and experimentally verified in~\cite{Schattschneider2014}. 

At  $z=z_B$, Eq.~\ref{eq:DLG} reads
\begin{equation}
\psi_{m,n}(r,\varphi,z_B) =LS_{m,n}(r,\varphi,z_B)\, e^{\ii\frac{k\,r^2}{2R(z)}} e^{-\ii(2n+|m|+1)\zeta(z_B)},
\end{equation}
where, 
\begin{eqnarray}
LS_{m,n} &=& \sqrt{\frac{n!}{\pi w_B^2 (n+|m|)!}} \left(\frac{r}{w_B}\right)^{|m|}\times \nonumber \\
&& L_n^{|m|}\left(\frac{ r^2}{w_B^2}\right)  e^{-r^2/2 w_B^2}  e^{i (m \varphi+k_z z_B)}.
\end{eqnarray}
It is not by chance that this is the wave function of a Landau state in cylindrical coordinates up to a phase factor that does not influence the rotation dynamics~\cite{Bliokh2012}.

The rotation dynamics  depends on the Larmor frequency, the OAM, the magnetic waist $w_B$, and the $z$-dependent vortex radius. It is therefore difficult to compare different experiments with each other and with theory and simulation. This problem can  be tackled by defining a Rayleigh frequency $\omega_R = \Omega w^2_B/w^2_0 = \hbar k / m_e z_R = v_z/z_R$ (this is the reciprocal time the electron takes to traverse the Rayleigh range of a diffracting LG mode) and a dimensionless variable along the optical axis $\zeta=z/z_R$.  
For diffracting LG modes, the rotation dynamics Eqs.~\ref{eq:analyticomega} and~\ref{eq:analyticomega1} expressed as a dimensionless rotation frequency follow a universal function (Lorentz function), 
\begin{equation}
\frac{\expectation{\omega(z)}-\sigma \Omega}{\sgn (m) \, \omega_R}=\frac{1}{1+\zeta^2}.
\label{eq:Dimensionless}
\end{equation}
True vortices will deviate from this  behavior. In any case, using Eq.~\ref{eq:Dimensionless} we can compare  experiments performed with different parameters, or numerical simulations with the analytical result for LG modes. 

\section{Simulation}
 
\subsection{Knife edge cutting}
\label{sec:KnifeEdgeCutting}

In order to gain access to the aforementioned peculiar rotational dynamics contained in the expectation value $\expectation{\omega(z)}$ we borrow a technique which was successfully applied in optics~\cite{Arlt20031573,Cui2012} as well as in electron physics~\cite{GuzzinatiPRL}. That is to break the circular symmetry of the annular shaped EVB using an electron blocking knife-edge (KE). This enables us to measure the azimuthal rotation angle $\varphi(z)$ of the truncated intensity pattern. Since $\omega=d\varphi/ dt = d\varphi/dz \, v_z$, where $dz/dt=v_z$ is the velocity of the electron along the $z$ axis, it is possible to map rotational frequencies onto the $z$-axis. Eq.~\ref{eq:OmegaXi} gives
\begin{equation}
\varphi(z)=\frac{\Omega}{v_z}\,\left(\sum_m m \, \int^z_{z_{df}} \expectation { \op{\xi}^{-2}}_m \, dz  +\int^z_{z_{df}} \sigma \, dz\right),
\label{eq:analyticphi}
\end{equation}
where $z_{df}$ is the defocus of the observation plane, see Sec.~\ref{sec:ExpResults}. Thus, spatial angular variations can be translated into rotational dynamics. 
For diffracting LG modes, Eq.~\ref{eq:DLG}, this can be expressed analytically by integrating Eq.~\ref{eq:analyticomega1} over $z$
\begin{eqnarray}
\varphi(z)& = &\frac{m}{|m|}\left(\arctan\left(\frac{z}{z_R}\right)-\arctan\left(\frac{z_{df}}{z_R}\right) \right) + \nonumber \\
& + & \frac{\Omega}{v_z} \, \sigma (z-z_{df}).
\label{eq:analyticphi2}
\end{eqnarray}

To clarify that this procedure does not significantly alter the measurement outcome, the influence of limiting the azimuthal range is studied in this section by looking at the Fourier series representation in the azimuthal angle.

An uncut EVB with quantized OAM $\hbar\,m$ can be written as
\begin{equation}
\psi_m=f(r) e^{i m \varphi}.
\label{eq:InitalState}
\end{equation}
with a radial amplitude $f(r)$.
When a sector of the vortex is blocked by a knife edge the resulting wave function $\psi_c$ can be expressed as a Fourier series in the azimuthal angle
\begin{equation}
\psi_c(r,\varphi)=f(r)\sum_\mu c_\mu e^{i (m+\mu) \varphi}
\end{equation}
with the same $r$-dependence as the inital state Eq~\ref{eq:InitalState}.
We can evaluate the expectation value Eq.~\ref{eq:ExpectValLzOverRsqrd} as we did above for the uncut vortex,
\begin{equation}
\expectation{\op{r}^{-2}\op{L}_z}=\braket{\psi_c | \hat L_z \hat r^{-2}| \psi_c}= \hbar\, \expectation { r^{-2}}_m \sum_\mu c_\mu^\ast c_\mu (m+\mu) 
\label{eq:ExpectValLzOverRsqrdCutVortex}
\end{equation}
where
\begin{equation}
\expectation { r^{-2}}_m =\int_0^\infty {|f(r)|^2 r^{-2}  \, r\, dr}.
\end{equation}
The Fourier coefficients are normalized
\begin{equation}
\sum_\mu c_\mu^\ast c_\mu =1
\end{equation}
and obey 
\begin{equation}
c_{-\mu}=c_\mu^\ast
\end{equation}
so that 
\begin{equation}
\sum_\mu c_\mu^\ast c_\mu (m+\mu)=m \sum_\mu c_\mu^\ast c_\mu = m.
\end{equation}
Thus, Eq.~\ref{eq:ExpectValLzOverRsqrdCutVortex} reduces to
\begin{equation}
\expectation{\op{r}^{-2}\op{L}_z}=\hbar  m \, \expectation { r^{-2}}_m 
\label{eq:ExpectValLzOverRsqrdCutVortexReduced}
\end{equation}
which is the same as Eq.~\ref{eq:ExpectValLzOverRsqrd} for the single uncut vortex in Eq.~\ref{eq:InitalState}.  

The conclusion is that cutting a sector of a single vortex does not change the rotation dynamics.

\subsection{Numerical simulation and error sources}
\label{sec:Numerical simulation and error sources}

Eq.~\ref{eq:OmegaXi} provides a simple way to obtain the angular velocity: given the OAM of a vortex, we  need the  moment  $\expectation{\xi^{-2}}$ as a function of $z$. In the plane of the cutting edge, this moment depends only on the radial density of the (cut) vortex, so we can use the standard FFT procedure based on the Fresnel propagator\footnote{The FFT results are known to be exact, except of the LR ($\sigma\Omega$-term in Eq.~\ref{eq:OmegaXi}), which can be added by using a so called co-rotating coordinate system~\cite{Reimer1984}.}.
One can even include the lens aberrations. So our approach was to calculate $\expectation{\xi^{-2}}$  for beam profiles as a function of the position $z$ of the obstructive edge, from which we quantify the rotation dynamics of the vortical structure via Eq.~\ref{eq:OmegaXi}.

A standard method to produce EVBs is to use holographic fork masks (a diffraction grating with a dislocation)~\cite{VerbeeckNature2010,McMorranScience2011}. It has been argued that EVBs created by this method carry OAM impurities~\cite{Clark2014}. The main reason are irregularities in the geometry of the mask. This means that a certain vortex order does not carry a quantized OAM any more. Although we have shown that the preparation of a sector of a given vortex does not change its rotation dynamics, see Eq.~\ref{eq:ExpectValLzOverRsqrdCutVortexReduced}, irregularities in the obstructing edge, as well as diffuse scattering at its rim may well cause OAM impurities in a vortex. 
Based on Eq.~\ref{eq:OmegaXi} for a vortex of the order $m_0$ we can make the ansatz
\begin{equation}
\expectation{ \omega(z)}= \Omega \left(m_0 \, \expectation{\op{\xi}^{-2}}_{m_0}+\sigma \right)+\Omega \sum_{m\neq m_0} m  \, \expectation{ \hat \xi^{-2}}_m.
\label{omegaimpurity}
\end{equation}
The last term in Eq.~\ref{omegaimpurity} causes a deviation from the rotation dynamics of the quantized vortex $m_0$. It is small because the coefficients measure the small admixtures of other OAMs, so $\expectation{\op {\xi}^{-2}}_m \ll \expectation { \op {\xi}^{-2}}_{m_0}$. 

Another potential source of errors is the diffraction at the obstructing edge. Instead of a rotated shadow image of the cut vortex, one observes a blurred half ring structure with Fresnel fringes. These fringes add to the challenge of measuring a rotation angle of the edge. Nevertheless, since these angles are measured in  a series of $z$-positions of the edge, the difference of two consecutive measurements is less sensitive to the  smoothly changing fringe contrast.

\section{Experimental}
\label{sec:ExpResults}

\begin{figure}[h]
\centering	\includegraphics[width=90mm]{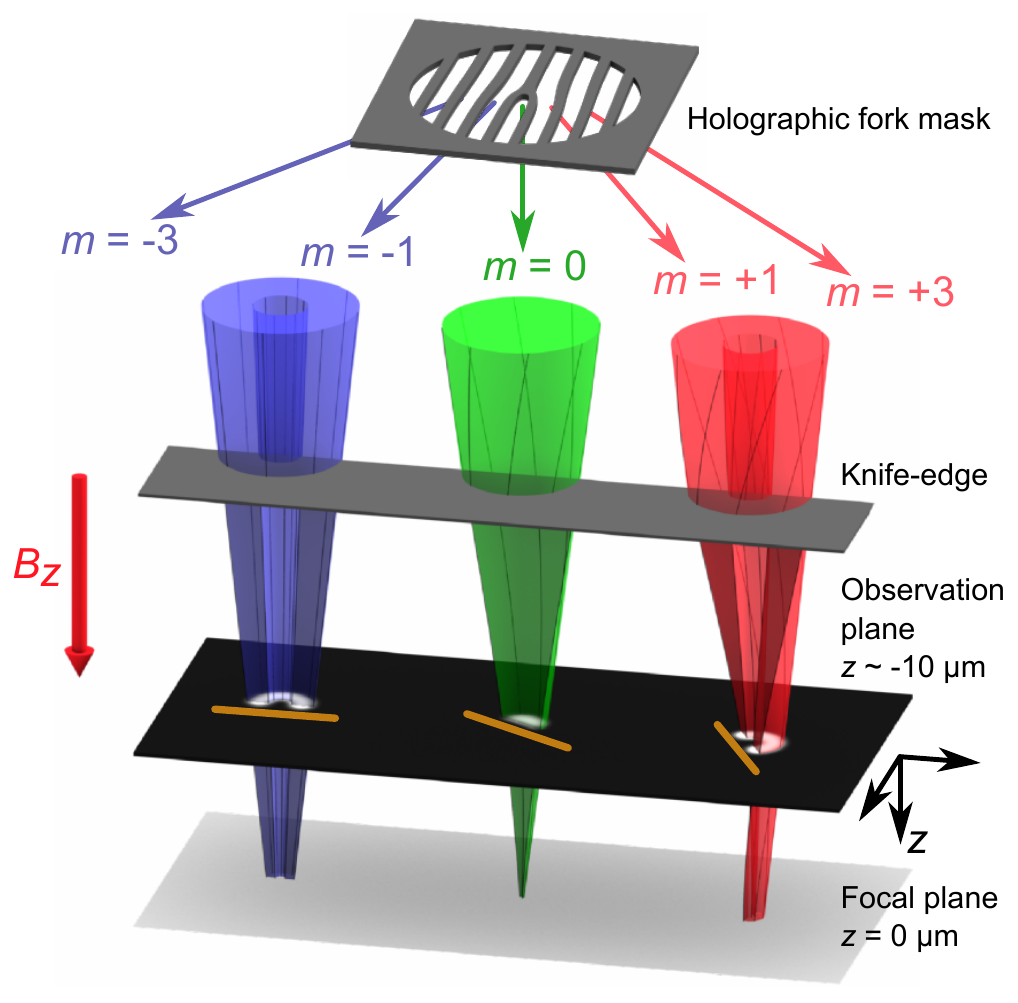}	
\caption{Schematics of the experimental setup, adopted from~\cite{Schattschneider2014}. A holographic fork mask splits the incident electron beam into EVBs such that the diffractions orders on either side of the zero beam carry negative and positive OAM ($m=...,-3,-1,0,1,3,...$), respectively. The convergent EVBs enter the longitudinal magnetic field $B_z$ of the objective lens and are incident on a knife-edge (KE). The cut EVBs propagate down the column reaching the observation plane at $z\approx\unit{-10}\micro\meter$, a few Rayleigh ranges above the focal plane. By varying the position of the KE the EVBs' rotational dynamics can be measured as azimuthal angle variation of the half moon like intensity patterns in the observation plane.}
\label{fig:ExpSetup}
\end{figure}

To test the peculiar rotational behavior of EVBs in magnetic fields predicted by Eq.~\ref{omega}, it is necessary to probe their internal azimuthal dynamics for a large range of different beam radii. To achieve that we placed a holographic fork mask in the C2 aperture holder of a FEI TECNAI F20 TEM working at \unit{200}{\kilo\volt}. By adjusting the C2 condenser strength, convergent EVBs can be produced in the KE plane, see Fig.~\ref{fig:ExpSetup}. These beams constantly change their radii when propagating along the $z$-direction. Their semi-convergence angle was $\unit{1.16}\milli\rad$. 
$B_z=\unit{1.9}\tesla$ for a  standard TEM objective lens and thus $w_B\sim\unit{25}\nano\meter$. With that, the radial extension of the EVBs can be expressed in units of $w_B$. It ranges from the radius of the used holographic aperture, in our case \unit{10.5}\micro\meter, equal to $\sim400 \, w_B$, down to the focused spot, with its characteristic dip in the center, showing a maximum intensity radius of \unit{0.9}\nano\meter, $\sim0.03\,w_B$, for $m=\pm1$ EVBs.

Due to the TEM geometry and the limited $z$-shift range of the specimen stage of $\pm\unit{375}\micro\meter$ the experimentally accessible range of different radii is reduced, ranging from $\sim20 \, w_B$ down to $\sim w_B/3$.

Note that the axial magnetic field within the accessible range of EVB radii can be considered as quasi-homogeneous showing deviations smaller than $\pm1.5\%$ of $B_z$. This has been tested in advance by investigating image rotations of copper grids and numerical simulations. The radial component of the magnetic lens field was calculated to be less than $10^{-6}B_z$.

\begin{figure}[h]
\centering	\includegraphics[width=90mm]{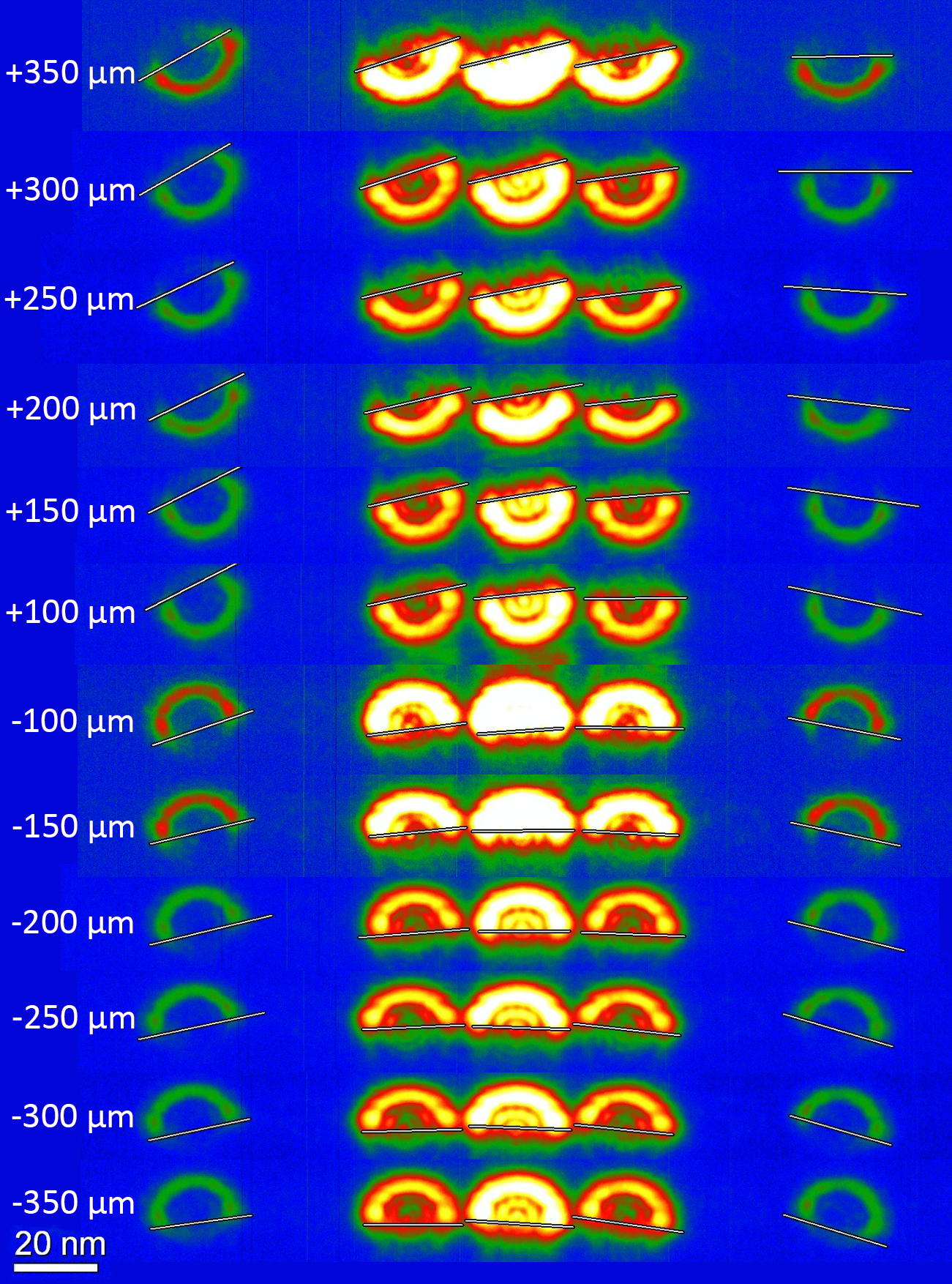}	
\caption{Experimental images of cut EVBs of a $z$-shift series. The non overlapping vortex orders $|m|=0,1,3$ are visible. The measured azimuthal rotation angle is inidcated as a faint solid line. Rotational dynamics can be observed by eye for all vortex orders.}
\label{fig:ImagesResultsLarmorNegm}
\end{figure}

In order to map the azimuthal dynamics of EVB onto the $z$-direction we obstruct half of the beam using a KE placed in the sample holder (see Sec.~\ref{sec:KnifeEdgeCutting}). As the Larmor frequency in the TEM objective lens field is of the order of $\Omega\sim\unit{2\pi \times19}\giga\hertz$, when using the relativistic electron mass $m_e=\gamma m_0$, the transverse rotational dynamics ($\Omega w_B \sim 10^{-5} c$) of EVBs are rather slow compared to the forward velocity of the relativistic electrons ($v_z\sim0.7c$). This leads to small but detectable pattern rotation of $\unit{3.3}\degree/\unit{100}\micro\meter$ when the KE is shifted up or down in the $z$-direction. To enhance the angular resolution the C2 condenser is under-focused (a few Rayleigh ranges $z_R\sim\unit{2}\micro\meter$) as seen from the observation plane (i.e. the rotated vortex is observed about $z_{df}\sim\pm \unit{10}\micro\meter$ from the focus, where the vortex orders do not overlap, see Fig.~\ref{fig:ImagesResultsLarmorNegm}).
To further increase the angular resolution the image contrast was enhanced using color coding and gamma correction.

Fig.~\ref{fig:ImagesResultsLarmorNegm} shows typical experimental images of the cut EVBs for the whole accessible $z$-shift range, including $|m|=0,1,3$ vortex orders. The angles between e.g. a horizontal line and the faint solid lines represent the measured azimuthal angles $\varphi(z)$ for different $z$-shift values of the KE, indicated next to the row of cut EVBs. By visual judgment alone one can already see rotational dynamics for all vortex orders. 

\begin{figure}[h]
\begin{flushleft}
\centering \includegraphics[width=90mm]{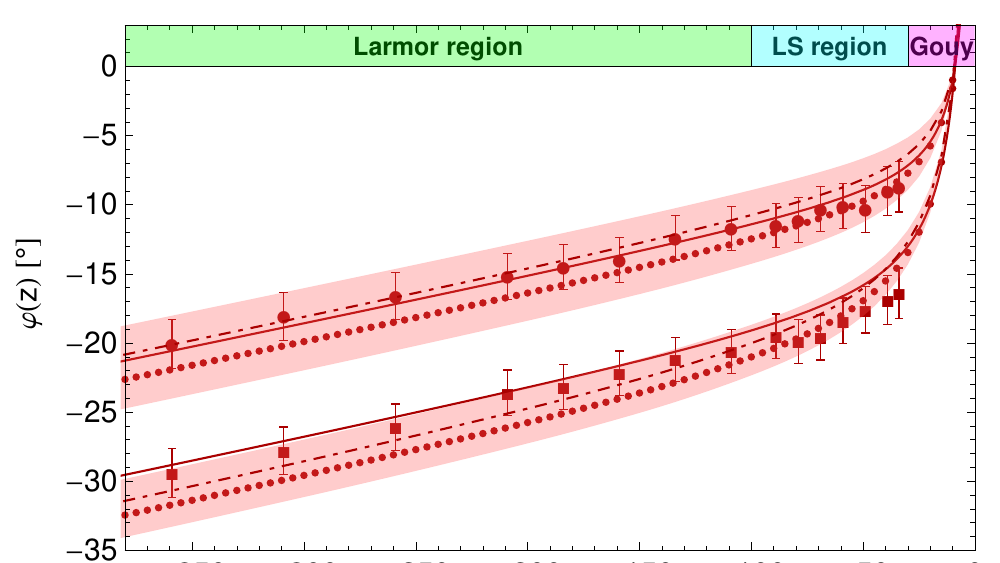}\\[-1pt]
\includegraphics[width=90mm]{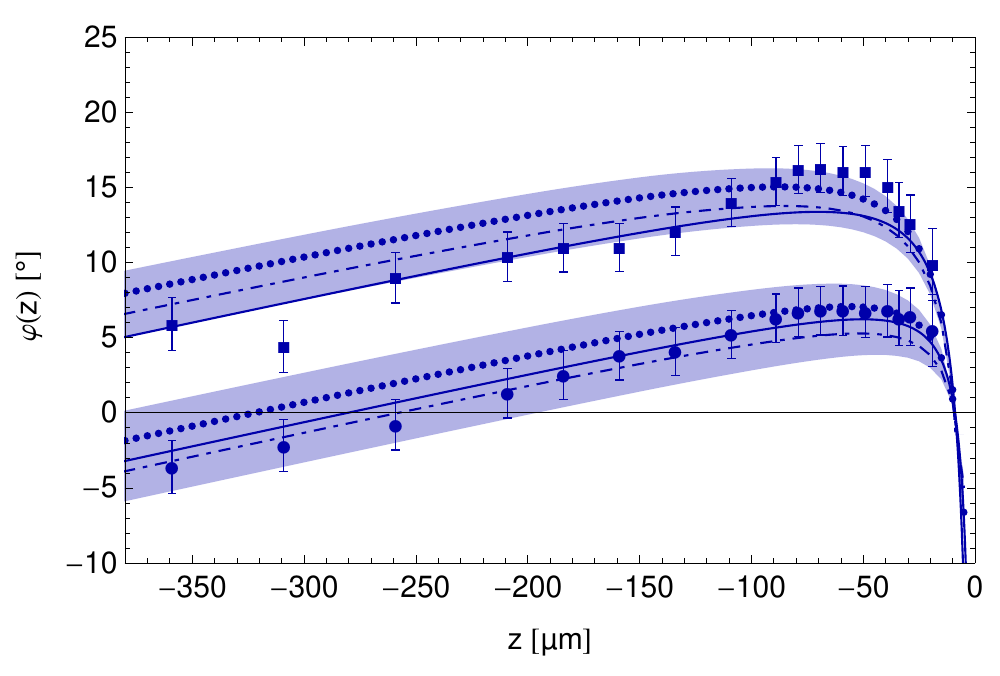}
\caption{Experimental data (large dots with error bars, $|m|=1$, squares $|m|=3$, upper diagram, red, $m>0$, lower diagram, blue, $m<0$) giving the azimuthal rotation angle $\varphi(z)$ of the cut EVBs over the $z$-shift value for an experiment scanning the LR- and  LS-region. Eq.~\ref{eq:analyticphi} was used to calculate the small dots using moments from the numerical simulation. For the solid lines, moments from the DLG modes, Eq.~\ref{eq:DLG}, were taken, using $z_R=\unit{1.46}\micro\meter$ for $|m|=1$ and $z_R=\unit{2.84}\micro\meter$ for $|m|=3$. The dot dashed lines show the influence of $35\%$ symmetric OAM impurities applied to the pure numerical simulate
d vortices using Eq.~\ref{omegaimpurity}, whereas the shaded areas indicate solely asymmetric OAM impurity contributions of $17.5\%$. Error bars include the estimated reading error, knife-edge roughness and stage positioning.}
\label{fig:ResultsLarmorLSNegPosm} 
\end{flushleft}
\end{figure}

To quantify these rotational dynamics, Fig.~\ref{fig:ResultsLarmorLSNegPosm} shows the measured azimuthal angles for LR- and LS-regions and compares them to the numerical simulated values as well as to the analytical model Eq.~\ref{eq:analyticphi2}. Finer $z$-scanning was used where the vortex size approaches the LS-size. This is also illustrated by the colored bar in the upper diagram in Fig.~\ref{fig:ResultsLarmorLSNegPosm} indicating the LR-region in green, which ranges from $z=\unit{-380}\micro\meter$ to $z\sim \unit{-100}\micro\meter$ and the LS-region in cyan, ranging from $z \sim \unit{-100}\micro\meter$ to $z \sim \unit{-40}\micro\meter$.
The region from $z \sim \unit{-30}\micro\meter$ to the focus $z=0$ represents the Gouy regime, in magenta. 

The measured azimuthal angles are in good agreement with the theoretical predictions. They show LR independent of the EVBs OAM for beam radii significantly larger than $w_B$. When the beam size approaches the LS-size, LR smoothly turns into quantized LS rotational behavior with no-rotation for negative OAM and cyclotron-rotation for positive OAM. When further decreasing the EVBs size well below $w_B$ the transition to fast Gouy rotation is observed, especially for negative OAM. Care must be taken that in particular for lower $z$-shift values diffraction effects and faint misalignments of the KE, meaning more or less cutting between the different $z$-shift values induce strong angular deviations. 

According to~\cite{Clark2014} the OAM admixtures for $m\neq m_0$ in Eq.~\ref{omegaimpurity} were estimated to $\unit{2.5} \%$ for $m = m_0 \pm 1$, $\unit{5} \%$ for $m = m_0 \pm 2$, $\unit{7.5} \%$ for $m = m_0 \pm 3$ and $\unit{2.5}\%$ for $m = m_0 \pm 4$. 
To account for the effect of these impurities on the rotation, these percentages were used in Eq.~\ref{omegaimpurity} to obtain upper/lower bounds for the deviation from the pure vortex behavior, indicated by colored bands in Fig.~\ref{fig:ResultsLarmorLSNegPosm}.
When applied symmetrically to the pure vortices, meaning $\unit{65} \% \, m_0$, $\unit{17.5} \%$ for all $m > m_0$ and $\unit{17.5} \%$ for all $m < m_0$, this results in a shift towards lower spreadings between the positive and negative vortex orders (see Fig.~\ref{fig:ResultsLarmorLSNegPosm}, dot dashed lines). This tendency can also be observed in the experimental data in Fig.~\ref{fig:ResultsLarmorLSNegPosm}, hinting at OAM impurity contributions.

The error estimates given in Fig.~\ref{fig:ResultsLarmorLSNegPosm} and Tab.~\ref{tab:RotFreq} are calculated as a combination of three effects: the first is the stage positioning error, which increases linearly for higher $z$-shift values. The relative stage error was given by the manufacturer to be of the order of $\unit{3} \%$. 
The second is the surface roughness of the used KE to block half of the incoming EVB. It turned out that the KE surface roughness is a crucial experimental parameter, because when the beam radii approach $w_B$, surface corrugations of the order of a few nanometers already introduce angular deviations of a few degrees. On the other hand for higher beam radii the surface roughness does not significantly contribute to the overall error. As a consequence of that, it is absolutely necessary to choose a very smooth KE with a surface roughness better than $R_z<\unit{1}\nano\meter$. We chose to take a $\expectation {111}$ Si-wafer that was broken along a low indexed zone axis, where $R_z$ was measured to lie below \unit{1}\nano\meter, thus keeping the contributions of the KE surface roughness well below $\unit{0.5}\degree$. 
The third contribution is the reading error stemming from the azimuthal cutting angle determination, which lies below $\unit{1.4}\degree$ in our case. 
Altogether the estimated error of the azimuthal angle determination techniques used is of the order of $\pm\unit{2}\degree$.

Tab.~\ref{tab:RotFreq} compares rotational frequencies in units of $\Omega$ averaged over the LR-region and the LS-region stemming from multiple experiments for the vortex orders $|m|=1,3$ with the numerical simulation and the diffracting LG model, Eq.~\ref{eq:analyticomega1}. It shows the remarkable agreement between the experimentally gained rotational frequencies and the theoretically expected ones, thus bringing further evidence that EVBs in magnetic fields exhibit peculiar rotations.

\begin{table}[!h]
\begin{center}
\begin{tabular}{l l |D{.}{.}{2.7} D{.}{.}{2.7}}
\multicolumn{2}{c|}{} & \multicolumn{1}{c}{LR-region} & \multicolumn{1}{c}{LS-region} \\
\hline
\multirow{3}{*}{$m=-3$} & DLG & 0.85\pm0.12 & -0.38\pm0.87 \\
& Numerical & 0.73\pm0.20 & -1.05\pm1.16 \\
& Experiment & 0.64\pm0.16 & 0.18\pm0.21 \\
\hline
\multirow{3}{*}{$m=-1$} & DLG & 0.89\pm0.08 & 0.03\pm0.61\\
&Numerical & 0.89\pm0.09 & 0.04\pm0.60 \\
& Experiment & 0.99\pm0.14 & 0.15\pm0.29 \\
\hline
\multirow{3}{*}{$m=+1$} & DLG & 1.11\pm0.08 & 1.97\pm0.61 \\
& Numerical & 1.11\pm0.09 & 1.96\pm0.60 \\
& Experiment & 1.25\pm0.14 & 1.85\pm0.27 \\
\hline
\multirow{3}{*}{$m=+3$} & DLG & 1.15\pm0.12 & 2.38\pm0.87 \\
& Numerical & 1.27\pm0.20 & 3.05\pm1.16 \\
& Experiment & 1.30\pm0.17 & 2.02\pm0.21 \\
\hline
\end{tabular}
\end{center}
\caption{Rotational frequencies in units of $\Omega$ for the numerical calculation, the diffracting LG approximation (Eq.~\ref{eq:analyticomega1}) and the experimental data. The errors indicated for the theory values represent standard deviations over the selected $z$-range, whereas the error for the experimental data gives the standard error of the mean weighted with the estimated experimental errors.
}
\label{tab:RotFreq}
\end{table}

Note that due to the averaging over extended $z$-shift regions and the smooth transition between different rotational regimes the rotational frequency of EVBs slightly increases or decreases compared to the LR in the LR region and to the zero- and cyclotron-rotation in the LS-region, respectively.

\begin{figure}[h]
\centering\includegraphics[width=90mm]{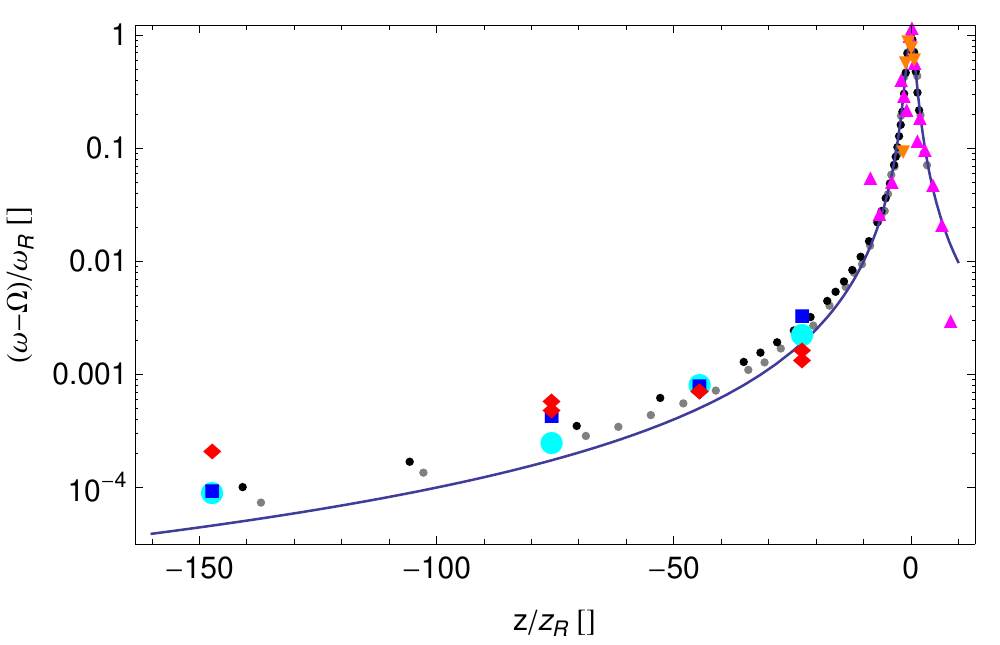}	
\caption{Comparison of various experimental results, theory and numerical simulations using the universal function Eq.~\ref{eq:Dimensionless}. The curve represents DLG modes, Eq.~\ref{eq:analyticomega1}, with the large cyan dots representing their averages in the LR- and LS-regions. The small gray and black dots mark the numerical simulation for $m=1,3$, respectively, with the blue squares indicating their averages in the respective regions. Red diamonds stand for the experimental values for $|m|=1,3$, averaged over the same ranges, whereas the triangles show experimental values for the Gouy-region. Magenta up triangle values where measured by the authors and the orange down triangles are taken from~\cite{GuzzinatiPRL}.} 
\label{fig:Dimensionless}
\end{figure}

The rotation dynamics included in Eq.~\ref{omega} represent our main findings, using the universal form Eq.~\ref{eq:Dimensionless}; it is possible to summarize experimental data, including Gouy-region data from the close vicinity of the focal plane, the numerical simulation and the diffracting LG model Eq.~\ref{eq:analyticomega1}, see Fig.~\ref{fig:Dimensionless}.
The logarithmic scale of Fig.~\ref{fig:Dimensionless} covers four orders of magnitude of the dimensionless rotation dynamics. It can be seen that the numerical simulations based on the moments of the defocused vortices follow the universal curve (valid for LG modes) up to $\zeta \sim 10$ quite well. For higher $z$-values, the $m=1$ simulation is slightly above the Lorentz curve, the $m=3$ simulation is well above. We have included the results of \cite{GuzzinatiPRL} taken with other parameters (voltage, convergence, method). The experimental results in the entire range covering Larmor, Landau and Gouy behavior are very close to the numerical simulations. In view of the experimental difficulties, and the rotation frequencies covering four orders of magnitude this is an extraordinary result. 

\section{Conclusions}	
The intuitive idea that the rotation dynamics of electron vortices can be described by a combination of slow Larmor and fast Gouy rotations  is illustrated by the present theoretical approach and confirmed by  experiments. Usual Larmor rotation is found for large vortex radii, far from the focus.  Rapid Gouy rotation appears close to the focus for narrow vortices.  Landau behavior with quantized rotation emerges as a special case when the vortices have radii close to the magnetic waist, bridging the Larmor and Gouy regimes.
The present quantum approach reconciles the three different regimes of rotational behavior (classical Larmor \cite{Glaser1952}, Landau \cite{Bliokh2012}, and rapid Gouy \cite{GuzzinatiPRL} rotation) of electron vortices in a magnetic field in a unifying description.

\section{Acknowledgements}
This work was supported by the Austrian Science Fund (FWF; grant no.
I543-N20).

\section*{References}

\bibliographystyle{unsrt} 
\bibliography{Lit2012}

\end{document}